\documentclass[10pt]{article}
\usepackage{geometry}
\usepackage{xcolor}
\usepackage{float}
\definecolor{DarkGreen}{RGB}{0, 100, 0}

\definecolor{graycolor}{gray}{0.9} 
\usepackage{microtype}
\usepackage{setspace} 
\usepackage[utf8]{inputenc}
\usepackage[english]{babel}
\usepackage{times}
\usepackage{array}
\usepackage{soul}
\usepackage{amsfonts,amsmath,amssymb}
\usepackage{latexsym,color,cite} 
\usepackage{cite}
\usepackage[numbers,sort&compress]{natbib}
\usepackage{natbib}

\usepackage{titlesec} 
\titleformat {\section} [block] {\raggedright \fontsize{10}{10}\selectfont\bfseries} {\thesection. \space} {0pt} {}
\titlespacing {\section} {0pt} {12pt} {6pt}
\titleformat {\subsection} [block] {\raggedright \fontsize{10}{10}\selectfont\itshape} {\thesubsection .\space} {0pt} {}
\titlespacing {\subsection} {0pt} {12pt} {6pt}
\titleformat {\subsubsection} [block] {\raggedright \fontsize{10}{10}\selectfont} {\thesubsubsection .\space} {0pt} {}
\titlespacing {\subsubsection} {0pt} {12pt} {6pt}
\titleformat {\paragraph} [block] {\raggedright \fontsize{10}{10}\selectfont} {} {0pt} {}
\titlespacing {\paragraph} {0pt} {12pt} {6pt}

\usepackage{array} \newcommand{\PreserveBackslash}[1]{\let\temp=\\#1\let\\=\temp}
\newcolumntype{C}[1]{>{\PreserveBackslash\centering}m{#1}}
\newcolumntype{R}[1]{>{\PreserveBackslash\raggedleft}m{#1}}
\newcolumntype{L}[1]{>{\PreserveBackslash\raggedright}m{#1}}
\usepackage{lineno}
\usepackage{tabularx}
\usepackage{colortbl}
\usepackage{graphicx}
\usepackage{float}
\usepackage[export]{adjustbox}
\usepackage{caption}
\usepackage{fancyhdr} 
\usepackage{lastpage}
\usepackage{layout}
\usepackage{setspace} 
\usepackage{enumitem}
\usepackage{booktabs}
\usepackage{arydshln}
\usepackage{multirow}
\usepackage{color}
\usepackage{hyperref} 
\hypersetup{
	colorlinks=true,
	linkcolor=blue,
	filecolor=blue,
	urlcolor=black,
	citecolor=cyan,
}

\setcitestyle{open={[},close={]},citesep={,\!},numbers}

\fancypagestyle{firstpage}{
    \setlength{\headsep}{2.2cm}
    
    \setlength{\footskip}{1.5cm}
    \fancyhf{}
    \lhead{\begin{table}[H]
        \centering
        \begin{tabular}{L{2.5cm}C{10cm}C{3.1cm}R{2cm}}
            \includegraphics[scale=0.035]{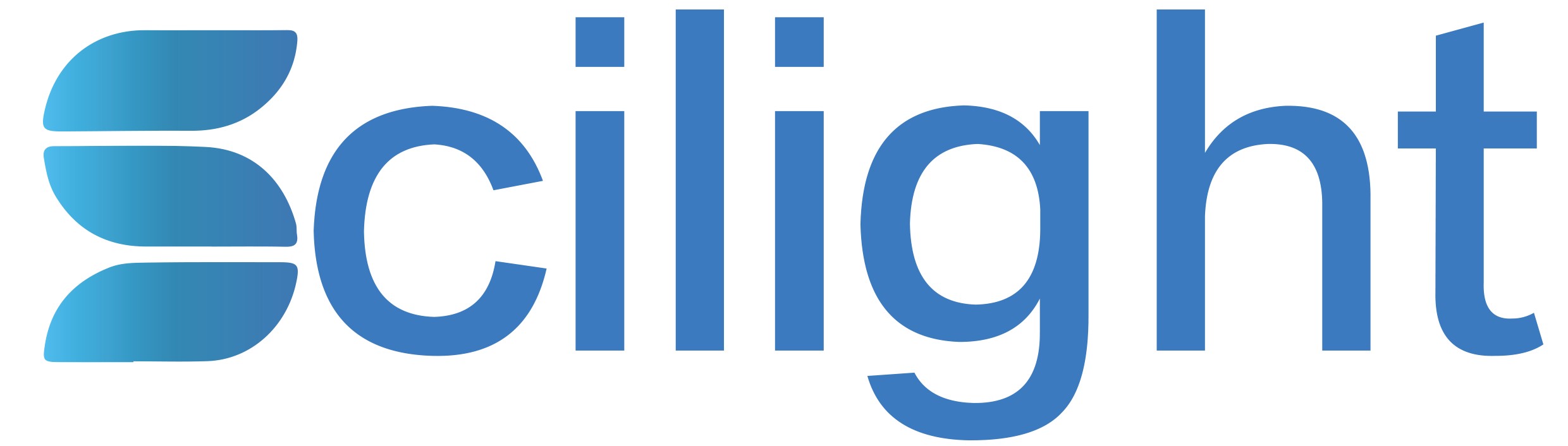} \vspace{-6pt}& \cellcolor{graycolor}\begin{tabular}[c]{@{}c@{}}\textit{International Journal of Gravitation and Theoretical Physics}\\ \href{https://www.sciltp.com/journals/ijgtp}{https://www.sciltp.com/journals/ijgtp}\end{tabular} & \includegraphics[scale=0.0150]{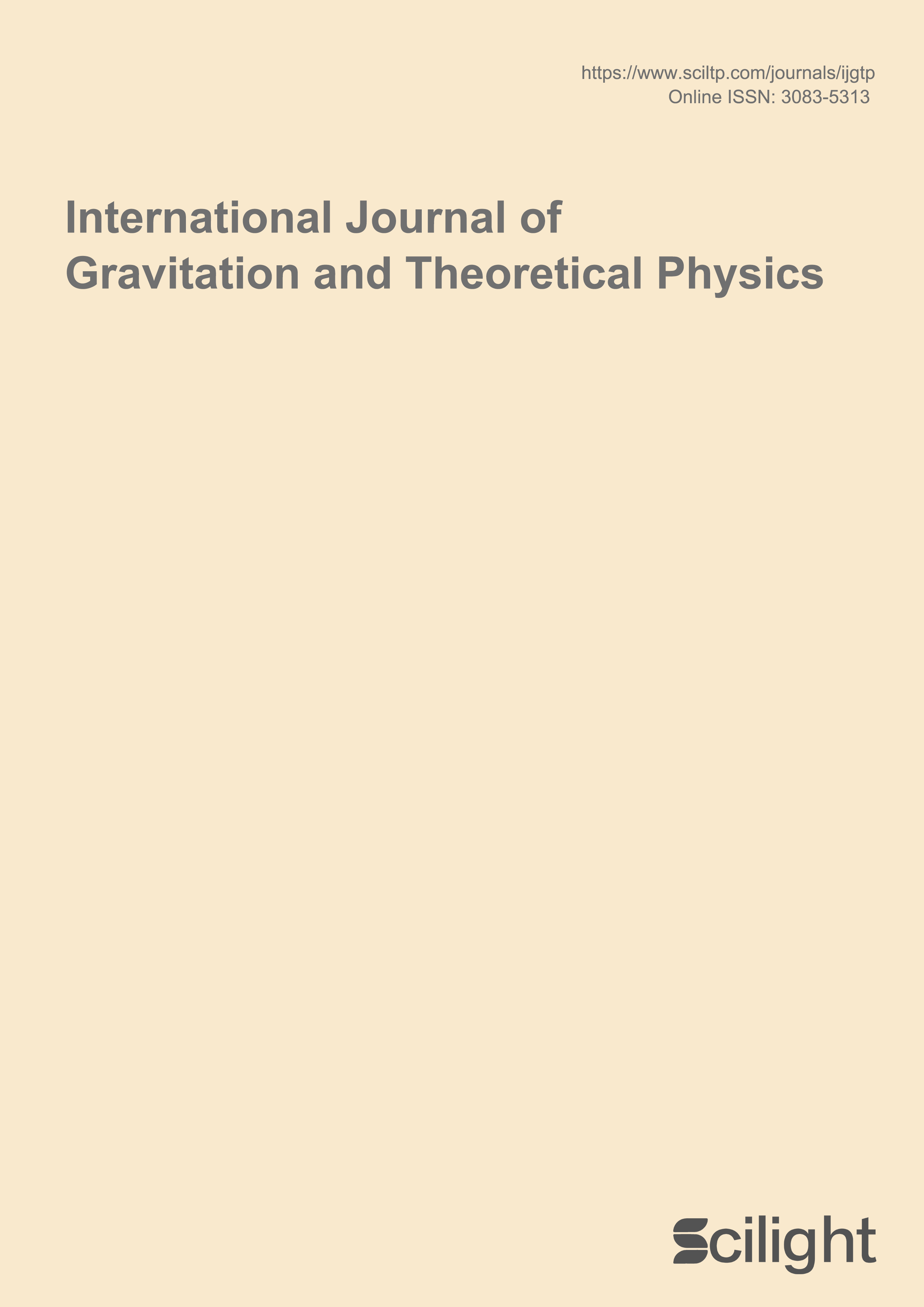} \vspace{-3pt}\\
        \end{tabular}
        \vspace{-22pt}
    \end{table}}
   
    \fancyfoot[C]{
        \vspace{-1.55cm}
        \begin{table}[H]
            \begin{minipage}[c]{0.15\columnwidth}
                \includegraphics[scale=0.5]{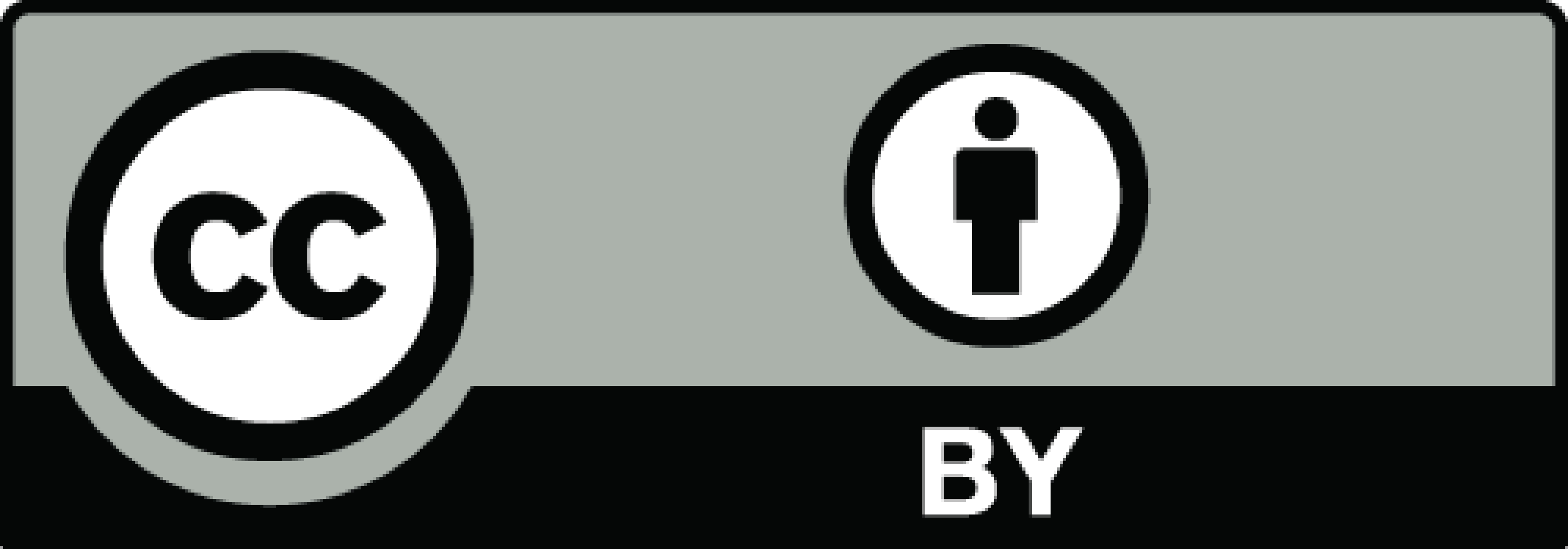} \vspace{1.1pt}
            \end{minipage}
            \hfill
            \begin{minipage}[c]{0.85\columnwidth}
                \scriptsize \textbf{Copyright:} © 2026 by the authors. This is an open access article under the terms and conditions of the Creative Commons Attribution (\mbox{CC BY}) license (\href{https://creativecommons.org/licenses/by/4.0/}{https://creativecommons.org/licenses/by/4.0/}). \\ \textbf{Publisher’s Note:} Scilight stays neutral with regard to jurisdictional claims in published maps and institutional affiliations.
            \end{minipage}
    \end{table}}
    \vspace{-0.55cm}
}

\begin{document}
\def\beq{\begin{equation}}
\def\eeq{\end{equation}}

\newcommand{\bear}{\begin{eqnarray}} 
\newcommand{\ear}{\end{eqnarray}}

\newcommand{\R}{{\mathbb R}}
\newcommand{\p}{\partial}
\newcommand{\nn}{\nonumber}

\def\R{{\mathbb R}}

\newgeometry{left=2.5cm, right=2.5cm, top=1.8cm, bottom=4cm}
	\thispagestyle{firstpage}
	\nolinenumbers
	{\noindent \textit{Article
    }}
	\vspace{4pt} \\
	{\fontsize{18pt}{10pt}\textbf{Photon sphere for a  dyonic black hole\\  corresponding to $A_2$ Toda chain}}
	\vspace{16pt} \\
	{\large Vladimir D. Ivashchuk \textsuperscript{1,2,*} and  Evgeny E. Trubach \textsuperscript{3}  }
	\vspace{6pt}
	 \begin{spacing}{0.9}
		{\noindent \small
			\textsuperscript{1}	Center of Gravitation and Fundamental Metrology, SRCAM Rostest, 
		Ozyornaya ul. 46, Moscow 119361, Russia \\
			\textsuperscript{2}	Institute of Gravitation and Cosmology, RUDN University, 
		ul. Miklukho-Maklaya 6, Moscow 117198, Russia \\
			\textsuperscript{3}	Department of Theoretical Nuclear Physics, National Research Nuclear University ``MEPhI'', 
		Kashirskoe sh. 31, Moscow 115409, Russia \\
        		    {*}  \parbox[t]{0.98\linewidth}{Correspondence: ivashchuk@mail.ru} 	\vspace{6pt}\\
		\footnotesize	\textbf{How To Cite}: Ivashchuk, V.D.; Trubach, E.E. Photon sphere for a  dyonic black hole\\  corresponding to $A_2$ Toda chain. \emph{International Journal of Gravitation and Theoretical Physics} \textbf{2026}, \emph{2}(3), 1. \href{https://doi.org/10.53941/ijgtp.2026.100015}{https://doi.org/10.53941/ijgtp.2026.100015}}\\
	\end{spacing}

\begin{table}[H]
\noindent\rule[0.15\baselineskip]{\textwidth}{0.5pt} 
\begin{tabular}{lp{12cm}}  
 \small 
  \begin{tabular}[t]{@{}l@{}} 
  \end{tabular} &
  \textbf{Abstract:} 
   We consider a dilatonic dyon black hole solution
 with a gravitational radius of  $2 \mu$ and two charges, 
 $Q_1$  (electric) and $Q_2$ (magnetic), within the framework 
 of a four-dimensional gravitational model comprising one scalar field and one 2-form. 
 The dilaton coupling constant $\lambda$ is fixed by $\lambda^2 = \frac{3}{2}$.
 This solution  is related to the $A_2$ Toda chain. 
 The circular orbits of null geodesics are explored, and the fifth-order polynomial 
 master equation governing the photon sphere radius $R_0$
 is analyzed. It is shown to admit a unique solution satisfying $R_0 > 2 \mu$. 
 The circular null geodesics are proven to be unstable. 
 Finally, the black hole shadow is examined, yielding relations 
 for the shadow angle  and the critical impact parameter.
\\
  & 
  \textbf{Keywords:} black hole; photon sphere;  dyon; Toda chain; black hole shadow  
\end{tabular}
\noindent\rule[0.15\baselineskip]{\textwidth}{0.5pt} 
\end{table}

\section{Introduction}

Interest in black holes, initially motivated by evidence for a supermassive black hole
at the center of our Galaxy \cite{Ghez}, has been further strengthened by the recent
detection of gravitational waves from a black hole merger \cite{Abbott}. 
Consequently, a large body of current literature is devoted to investigating 
black holes in modified theories of gravity (see \cite{Vagn} and references therein). 
In this work, we consider one such theory - namely, a dilatonic scalar-tensor model comprising 
a single scalar field and one Abelian gauge field.
 
 Here we deal with photon spheres which control several observable and theoretical features of black holes. In the eikonal limit the real part of the quasinormal-mode frequency is set by the orbital frequency of the unstable circular null geodesic, while the imaginary part is set by the Lyapunov exponent that measures the instability of that orbit \cite{CMBWZ,KZh}. The same geodesic determines the critical impact parameter and therefore the silhouette of the black hole as seen by a distant observer; the resulting shadow radius is now constrained by Event Horizon Telescope images of $Sgr A^*$ and $M87^*$
 \cite{Vagn,PTs}. Photon spheres also govern the high-frequency absorption cross-section and, for a radiation fluid, coincide with Bondi’s sonic horizon \cite{CGP}. These applications have stimulated a systematic study of circular null geodesics in modified-gravity and supergravity black holes. 
   
This paper builds upon previous research \cite{ABDI,ABI,BBIM,MBI} 
devoted to dilatonic dyonic and dyon-like black hole solutions. 
The present results may be regarded as a modest contribution to 
the extensive body of work addressing spherically symmetric configurations, 
including black holes and black branes, as reviewed in \cite{BronShikin}--\cite{Ifbb}
 and the references therein. Such solutions naturally arise in gravitational models 
 that incorporate scalar fields and antisymmetric forms.

Here we consider a dilatonic dyon black hole solution \cite{ABDI} with electric and magnetic charges 
 $Q_1$ and $Q_2$, respectively, within the framework of a four-dimensional model comprising a metric 
$g$,  a scalar field $\varphi$, and a 2-form $F$. 
The corresponding dilatonic coupling constant is denoted by $\lambda$.
Here we put
\beq \label{i4.1}
  \lambda  = \pm \sqrt{3/2}.
 \eeq    
  and deal with a dyon configuration on an oriented manifold
  ${\cal M } = (2\mu, + \infty)  \times S^2 \times  \R $ ($\mu > 0$), 
  with the 2-form field 
  \beq
  F = Q_1 f * \tau + Q_2 \tau, \label{0.1}
  \eeq
  where $\tau = {\rm vol}[S^2]$  is the volume form on the sphere, $* = *[g]$
  denotes the Hodge operator associated with $({\cal M }, g)$, 
  and $f$  is a  scalar function on ${\cal M }$.
 
   In Ref. \cite{ABI}, the term ``dyon-like'' configuration was introduced 
   to distinguish the non-composite ansatz 
    \beq
   F^{(1)} = Q_1 f * \tau,  \qquad F^{(2)} = Q_2 \tau, \label{0.1nc}
   \eeq
   where $F^{(s)}$, $s =1, 2$, are two 2-forms,   from the composite dyonic one 
   given in (\ref{0.1}). 
     
In previous studies, specific dyon-like solutions for two Abelian 2-form fields were investigated for a range of Lie algebras and dilatonic couplings \(\lambda_1, \lambda_2\) (see Refs. \cite{GM, KLOPP, ABI, Dav, GalZad, AIMT}). The particular case of the \(A_1 + A_1\) Lie algebra, which corresponds to the condition \(\lambda_1 \lambda_2 = \tfrac{1}{2}\), was examined in Refs. \cite{GM, KLOPP, ABI, AIMT, IMNT}. When the couplings are equal, \(\lambda_1 = \lambda_2 = \lambda\), one is led to relation  $ \lambda  = \pm \sqrt{1/2}$ , which appears in certain four-dimensional supergravity and string-induced models \cite{GM,KLOPP,CGP}. Dyonic (composite) solutions have also attracted considerable attention, as referenced in \cite{ChHsuL,GKLTT,PTW,Br0,ABDI,IKMN} and related works. In fact, the dyonic solution from Ref. \cite{ChHsuL} (see also \cite{Br0}) is a non-composite version of the solutions originally found in Refs. \cite{GM,KLOPP}.
 
This paper examines circular null geodesics that exist at a fixed radius 
$R = R_0$, 
which is equivalent to the photon-sphere radius. The characteristics of photon spheres 
play a key role in numerous studies involving black hole solutions, 
ranging from the eikonal-limit spectra of quasinormal modes (QNM) and 
black hole shadow radii to circular orbits for massive test particles, 
among other phenomena (see Refs. \cite{Vagn,KZh, CGP,BTIMNU, PTs} and the bibliography therein).

   Photon spheres of static spherically symmetric supergravity black holes were analysed by Cveti\v{c}, Gibbons and Pope 
   \cite{CGP}. In particular, in subsection 3.4 they considered (for zero gauge coupling $g$) the same non-extremal Einstein–Maxwell–dilaton dyon black hole (with dilaton coupling equivalent to \(\lambda^2=3/2\)) in the Lü–Pang–Pope parametrization \cite{LPP} of the moduli functions $H_s$ (by fluxbrane polynomials \cite{Ifbb} depending upon  $f_0 = 1 - \frac{2\mu}{R}$), and proved that the first-order (differential) condition for a circular null geodesic admits exactly one root outside the horizon which is necessarily the radius of (unstable) photon sphere. The present paper treats the same   non-extremal $A_2$ Toda dyon black hole in  parametrization of moduli functions \(H_s=1+P_s/R+P_s^{(2)}/R^2\) (by black brane polynomials \cite{IMp3}). We derive the explicit fifth-order master polynomial, give an elementary uniqueness proof based on the strict concavity of an auxiliary polynomial $Q(z)$, prove instability by a direct computation of the second
   derivative of the effective potential $U''(R_0)$, obtain several exact roots and a table of numerical values, and compute the shadow observables for this metric.
   
A companion analysis for the complementary coupling \(\lambda^2=1/2\) with moduli functions \(H_s=1+P_s/R\) and cubic master equation has recently appeared in Ref. \cite{IKMN} devoted to photon sphere for non-extremal $A_1+A_1$ Toda dyon black hole.
The two papers therefore treat two distinct members of the same family with
different algebraic degrees of the master equations and the explicit formulae.

The organization of the present paper is as follows. The gravitational setup and 
the explicit dyon black-hole solution are introduced in Section 2, 
while a compilation of its key physical parameters is provided in Section 3. 
Section 4 contains the core dynamical analysis: we derive the governing 
master equation for the null circular orbit radius $R_0$, 
rigorously prove the uniqueness of its root beyond the horizon 
$R_0 >  2 \mu$  and verify the instability of these photon orbits. 
 Subsequent to this, Section 5 supplies several illustrative exact solutions 
 to the derived polynomial equation. We close the analytical treatment 
 in Section 6 with study of the black hole shadow.

\section{Black hole dyon solution}

Let us consider a model governed by the action
\begin{eqnarray}
 S= \frac{1}{16 \pi G}  \int d^4 x \sqrt{|g|}\biggl\{ R[g] -
   g^{\mu \nu} \partial_{\mu} \varphi  \partial_{\nu} \varphi
   - \frac{1}{2} e^{2 \lambda \varphi} F_{\mu \nu} F^{\mu \nu }
 \biggr\},   \label{i.1} 
\end{eqnarray}
where $g= g_{\mu \nu}(x)dx^{\mu} \otimes dx^{\nu}$ is  metric,
 $\varphi $ is the  scalar field, 
 $F = dA  =  \frac{1}{2} F_{\mu \nu} dx^{\mu} \wedge dx^{\nu}$
is the $2$-form with $A = A_{\mu} dx^{\mu}$; 
$G$ is the gravitational constant;
 $\lambda$ is  coupling constant  
 obeying (\ref{i4.1})  and  $|g| =   |\det (g_{\mu \nu})|$.
 
 Here  we deal with the case
 \beq \label{ii.2}
 \lambda^2 = \frac{3}{2}
 \eeq
 which corresponds to Kaluza-Klein reduction from $5D$ gravitational model \cite{Lee}. 
   
 We consider  dyon black hole solution \cite{ABDI} to the field equations corresponding to the action
(\ref{i.1})  which is defined on the manifold
\beq \label{i.2}
 {\cal M }  =    (2\mu, + \infty)  \times S^2 \times  \R,
\eeq
and has the following form
\bear  
 ds^2 = (H_1 H_2 )^{1/2}
 \biggl\{ -  H_1^{-1 } H_2^{-1} 
 \left( 1 - \frac{2\mu}{R} \right)  dt^2  
  +  \frac{dR^2}{1 - \frac{2\mu}{R}} + R^2  d \Omega^2_{2}
  \biggr\},  \label{i.3}  \\  
 \exp(\varphi)=  H_1^{\lambda/2 } H_2^{- \lambda/2}, \label{i.3a}
 \\  \label{i.3bem}
 F=  \frac{Q_1}{R^2}   H_{1}^{-2} H_2  dt \wedge dR
     +  Q_2 \tau.
\ear

Here  $Q_1$ is electric charge and $Q_2$ is magnetic charge, 
$\mu > 0$,  $d \Omega^2_{2} = d \theta^2 + \sin^2 \theta d \phi^2$
is the  metric on the unit sphere $S^2$
 ($0< \theta < \pi$, $0< \phi < 2 \pi$) and
 $\tau = \sin \theta d \theta \wedge d \phi$
is the  volume form on $S^2$. 
For the speed of light we put $c=1$.

The functions $H_s$ are given by
\beq \label{i4.5}
H_s = 1 + \frac{P_s}{R} + \frac{P_s^{(2)}}{R^2},
\eeq
where
\bear \label{i4.6}
 2 Q_s^2 = \frac{P_s (P_s + 2 \mu) (P_s + 4 \mu)}{P_1 + P_2 + 4 \mu},
 \\ \label{i4.6a}
 P_s^{(2)} = \frac{P_1 P_{2} (P_s + 2 \mu) }{2 (P_1 + P_2 + 4 \mu)},
\ear
$s = 1,2$.

The (moduli) functions $H_s$ are generated by a Toda chain associated with the Lie algebra 
 $A_2$ \cite{ABDI}. (For a pioneering spherically symmetric dyon soluton governed by 
 $A_2$ Toda chain see Ref. \cite{Lee}.)
They obey the following boundary conditions:
\beq \label{i3.1a}
  H_s  \to H_{s0} = 1 + \frac{P_s}{2 \mu} + \frac{P_s^{(2)}}{(2 \mu)^2}  > 0
\eeq
for $R \to 2\mu $, and
\beq \label{i3.1b}
  H_s    \to 1
\eeq
for $R \to +\infty$, $s = 1,2$.

The solution is  a well-defined one   for $R > 2\mu$.
It is a composite brane solution  describing a configuration 
of two intersecting non-extremal black $0$-branes (electric one 
and a magnetic one) in case when  the intersection rule corresponds 
to the Lie algebra $A_2$ ($A_2 = sl(3)$) \cite{IMtop,Isym}.

Due to the first boundary condition (\ref{i3.1a}), we are led to a horizon at 
$R =  2 \mu$ while the second condition 
(\ref{i3.1b}) implies the asymptotic flatness of the metric as 
$R \to +\infty$.

For a global extension of the metric (\ref{i.3}), initially defined for 
 $R > 2 \mu$, we get  two horizons at $R = 2 \mu$ 
 and $R = 0$,  and a singularity at certain 
 $R_{*} < 0$.  ( Here $R_{*} < 0$ is the maximal (negative) root of the equation $ H_1(R) = 0$ for $P_1 > P_2$
  and it is maximal (negative)  root of another  quadratic equation  $ H_2(R) = 0$ for $P_1 < P_2$. 
  For $P_1 = P_2 = P > 0$ we have $R_{*} = - \frac{P}{2}$ which obeys both equations $H_s(R) = 0$, $s =1,2$.)  
  
  In a special case when   $P_1 = P_2 = P$ the  metric (\ref{i.3}) 
 coincides with the  Reissner-Nordstr\"om metric. 
  Indeed, in this case  we get 
 \beq \label{i.RN}
 ds^2 = H
  \biggl\{ -  H^{-2 } 
  \left( 1 - \frac{2\mu}{R} \right)  dt^2    +  \frac{dR^2}{1 - \frac{2\mu}{R}} + R^2  d \Omega^2_{2}
   \biggr\},  
  \eeq  
where $H = H_1 = H_2 = \left( 1 + \frac{P}{2R} \right)^2$  and $\varphi = 0$. 
By changing the radial variable $r = R +  \frac{P}{2}$ we obtain
  \beq  
   ds^2 = - A(r)   dt^2    +  \frac{dr^2}{A(r)} + r^2  d \Omega^2_{2}
      \label{i.RN_true}
    \eeq  
  where
     \bear \label{i.RN_A}
     A(r) = 1 - \frac{2GM}{r} +  \frac{Q^2_{RN}}{r^2},  \\
   2GM = P + 2\mu, \qquad Q^2_{RN} = \frac{1}{4} P (P+ 4 \mu) =  Q^2_{1}= Q^2_{2}.
   \label{i.RN_M_Q}
   \ear

\section{Physical parameters}

Here, we outline some physical parameters associated with the black hole solution \cite{ABDI}.

\subsection{Gravitational mass and scalar charge}
 
The ADM gravitational mass $M$ may be found from (\ref{i.3}) as
  \beq \label{i5.1}
 GM =   \mu +  \frac{1}{4} (P_1 + P_2),
\eeq
where $G$ is the gravitational constant.
 
For the scalar charge we find from (\ref{i.3a})
\beq \label{i5.1s}
 Q_{\varphi} =  \frac{1}{2} \lambda (P_1 -   P_2).
\eeq
 
By using relations (\ref{ii.2}), (\ref{i5.1}), and (\ref{i5.1s}), 
we obtain the following identity \cite{ABDI}
 
 \beq \label{i5.1id}
     2 (GM)^2   +     Q_{\varphi}^2   = Q_1^2 + Q_2^2 + 2 \mu^2.
 \eeq
(For the extremal case $\mu = +0$ see also Ref. \cite{PTW}.)

\subsection{The Hawking temperature and  entropy}

  The Hawking temperature, calculated from the metric (\ref{i.3}), 
  is  \cite{ABDI}:
    \beq \label{i5.2}
  T_H=   \frac{1}{8 \pi \mu} (H_{10} H_{20})^{-1/2}.
  \eeq
Here $H_{s0}$ are defined in (\ref{i3.1a}) and 
 relations $c = \hbar = k_B = 1$ are adopted.

For the Bekenstein-Hawking (area) entropy $S = A/(4G)$ ($A$ is the horizon area),
corresponding to the horizon at $R = 2\mu$, we obtain
 \beq \label{i5.2s}
 S_{BH} =   \frac{4 \pi \mu^2}{G}  (H_{10} H_{20})^{1/2}.
 \eeq
We note that relations (\ref{i5.2}) and (\ref{i5.2s}) imply 
 $T_H  S_{BH} =   \frac{\mu}{2G}$.

\section{Circular geodesic solutions}

In this section we study the geodesic equations corresponding to the metric (\ref{i.3}),
 restricting our attention to circular solutions.
 
\subsection{Geodesic equations}

The geodesic equations are equivalent to the Euler-Lagrange equations
\beq \label{4.ELeqs}
    \frac{d}{d\tau}\left(\frac{\partial \mathcal{L}}{\partial \dot{x}^{\alpha}}\right)
     - \frac{\partial \mathcal{L}}{\partial x^{\alpha}} = 0.
\eeq
following from the Lagrangian
\beq \label{4.Lag}
    \mathcal{L} = \frac{1}{2}  g_{\alpha \beta}(x) \dot{x}^{\alpha}\dot{x}^{\beta}.
\eeq

Here $\dot{x}^{\alpha}=dx^{\alpha}/d\tau=u^{\alpha}$ denotes the 4-velocity vector, with 
 $x^{\alpha}=x^{\alpha}(\tau)$ ($\alpha=0,1,2,3$). The parameter $\tau$ represents the proper time for a massive point particle moving along a timelike geodesic, and serves as the affine parameter in the null case. We adopt the normalization
 \beq \label{4.normal}
      g_{\alpha \beta}(x) u^{\alpha} u^{\beta} = -k = 2 {\cal E},
 \eeq
 where $k= 0,1$ corresponds to  null and timelike geodesics, respectively. 
The quantity ${\cal E}$ is the conserved energy integral obtained from the Lagrange equations for the Lagrangian (\ref{4.Lag}). 

In what follows we use the relations for  the redshift function $A(R)$ and central function $C(R)$: 
\bear
A =A(R) = (H_1 H_2)^{-1/2} \left(1-\frac{2\mu}{R}\right),
    \label{4.A} \\   
C =C(R) = (H_1 H_2)^{1/2} R^2.      \label{4.C}
\ear

In this subsection, we restrict our analysis to null and time-like geodesics confined to the equatorial plane:
 $\theta = \pi/2$. The symmetry then reduces the system to an effective three-dimensional problem, 
 for which the Lagrangian, derived from the metric in Eq. (\ref{i.3}), is given by
\beq \label{4.Lagrangian}
     \mathcal{L}_{*}
    = - A(R) \dot{t}^2 + (A(R))^{-1} \dot{R}^2 + C(R) \dot{\phi}^2.
\eeq

The Lagrangian $\mathcal{L}_{*}$
governs the geodesic equations which correspond to the variables  $t$, $R$ and $\phi$. 
The geodesic equation for the  variable $\theta$ is satisfied identically 
due to our choice of $\theta = \pi/2$.

For the cyclic coordinates $t$ and $\phi$ we find the following integrals of motion:
\beq \label{4.ELhat}
  \hat{E} = A(R) \dot{t},\qquad  \hat{L} = C(R) \dot{\phi}.
\eeq
For $k=1$, they describe the total energy $E = \hat{E} m$ 
and angular momentum $L = \hat{L} m$ of a test (neutral point-like) particle with mass $m > 0$.

Equation (\ref{4.normal}) takes the following form for the metric  (\ref{i.3})
\beq \label{eq:k}
    -  A(R) \dot{t}^2 + 
       (A(R))^{-1} \dot{R}^2 + C(R) \dot{\phi}^2 = -k. 
\eeq
By using (\ref{4.ELhat}), we are led to relation  
\beq \label{4.EqWithEL}
    -\frac{ \hat{E}^2}{A} + 
    \frac{ \dot{R}^2}{A} + \frac{\hat{L}^2}{ C} = -k,
\eeq
which can be rewritten as follows
\beq \label{4.EqForR}
    \dot{R}^2 + U(R) = \hat{E}^2.
\eeq
Here
\bear \label{4.eff_pot}
    U = U(R) = A(R) \left( k +  \frac{\hat{L}^2}{ C(R) }   \right) \qquad  \\ \nonumber
    = (H_1 H_2)^{-1} \left(1-\frac{2\mu}{R}\right)
    \left( k (H_1 H_2)^{1/2}  + \frac{\hat{L}^2}{ R^2}\right)
\ear
is the effective potential.

  The Lagrange equation for the radial coordinate $R$ reads
 \beq \label{4.EqForRtrue}
   2 \ddot{R} +  \frac{d U}{d R} = 0.
\eeq

\subsection{Circular photon orbits: non-extremal case}

\begin{figure}
  \includegraphics[width=\columnwidth]{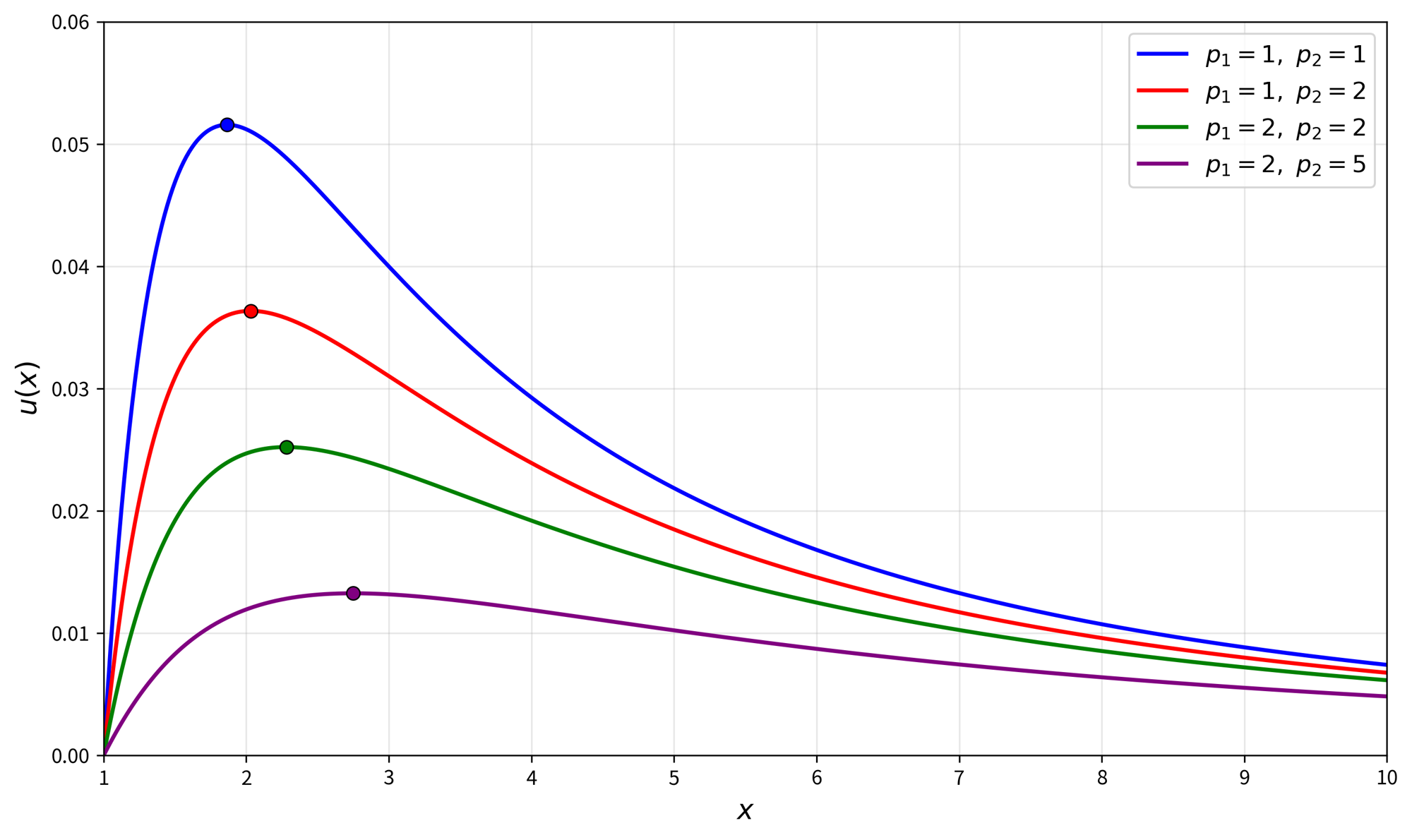}
  \caption{ The reduced effective potential 
  $u(x)$ from \eqref{4.u} as a function of $x$  for  particular sets of the parameters $p_1, p_2$.}
  \label{fig:u_x}
  \end{figure}

Now we consider null geodesics. For the effective potential we get
\bear \label{4.eff_pot_1}
    U = A(R) \frac{\hat{L}^2}{ C(R) } = (H_1 H_2)^{-1} \left(1-\frac{2\mu}{R}\right)
     \frac{\hat{L}^2}{ R^2},
\ear
or
\begin{equation} \label{4.eff_pot_x}
    U =  \frac{\hat{L}^2}{(2 \mu)^2 } u,
\end{equation}
where
\begin{equation} \label{4.u}
    u = u(x) = \frac{x -1}{ H_1(x) H_2(x) x^3},
\end{equation}
and 
\begin{equation} 
  \label{4.xp}
  x = R/(2 \mu), \qquad p_s = P_s/(2 \mu),
\end{equation}
are  dimensionless parameters, $s = 1,2$. See Figure \ref{fig:u_x}.
Here 
\beq \label{i4.5x}
H_s(x) = 1 + \frac{p_s}{x} + \frac{q_s}{x^2},
\eeq
where
\begin{equation} 
  q_s = p_s^{(2)} = \frac{p_1 p_2 (p_s + 1)}{2 (p_1 + p_2 + 2)},
\label{i4.qs}
\end{equation}
$s = 1,2$.

   Let us consider circular solutions for the null geodesic equations obeying
\beq \label{4.R0}
  R (\tau) = R_0 = {\rm const}, \qquad R_0 > 2 \mu.
 \eeq 
 Here $R_0$ is the radius of the photon sphere which  is covered by  circular photon orbits.

  In this case $\hat{L} \neq 0$ and the radial equation (\ref{4.EqForRtrue}) for 
 $R = R_0$  with $k= 0$ reads 
  \begin{equation}           
   \frac{dU}{dR} = 0, 
    \label{4.masteqR_0}
  \end{equation} 
where
\begin{equation}           
      \frac{dU}{dR} = \frac{1}{2 \mu}  \frac{dU}{dx} =   \frac{\hat{L}^2}{(2 \mu)^3 } 
      \frac{d u}{dx}.  
       \label{4.dUdR}
\end{equation} 
The calculations gives us 
\begin{equation}           
       \frac{d u}{dx} = -  \frac{ F(x)}{4 (p_1 + p_2 + 2)^2 x^8(H_1 H_2)^2},  
       \label{4.dudx}
\end{equation} 
where
  \begin{eqnarray}   
     F(x) = 8(p_1 + p_2 + 2)^2 x^5  
   +   4(p_1 + p_2 - 3)(p_1 + p_2 + 2)^2 x^4 \nonumber \\ 
       - 8(p_1 + p_2)(p_1 + p_2 + 2)^2 x^3   
     -4p_1 p_2(p_1+p_2+ 2)(p_1 p_2+2 p_1+2 p_2+3)x^2  \nonumber \\ 
        - 2p_1^2 p_2^2(p_1 + 1)(p_2 + 1)x  
    +  p_1^2 p_2^2(p_1 + 1)(p_2 + 1). \qquad \qquad
        \label{4.F}
 \end{eqnarray}  
The extremality condition  (\ref{4.masteqR_0}) is equivalent
to the master equation
\begin{equation}
F(x) = 0.
\label{4.masteqx}
\end{equation}

For future applications it is worth to rewrite dimensionless master equation 
(\ref{4.masteqx})  in terms of radial variable $R= x (2 \mu)$, where $\mu >0$. 
By multiplying (\ref{4.masteqx}) on $(2 \mu)^7$ we obtain
\begin{eqnarray}   
     8(P_1 + P_2 + 4 \mu)^2 R^5  
   +   4(P_1 + P_2 - 6 \mu )(P_1 + P_2 + 4 \mu )^2 R^4 \nonumber \\ 
       - 16 \mu  (P_1 + P_2)(P_1 + P_2 + 4 \mu )^2 R^3  
     - 4 P_1 P_2(P_1+P_2+ 4 \mu)(P_1 P_2+ 4 (P_1+ P_2) \mu +12 \mu^2 )R^2  \nonumber \\ 
        - 2 P_1^2 P_2^2(P_1 + 2 \mu)(P_2 + 2 \mu)R \quad 
    +  2 \mu P_1^2 P_2^2(P_1 + 2 \mu)(P_2 + 2 \mu) = 0. \qquad \qquad
        \label{B.FRmaster}
 \end{eqnarray}  
We remind that $P_s = 2 \mu p_s$, $s =1,2$. 

We note that relation (\ref{4.dudx}) just follows from
identity 
\begin{equation}           
       \frac{d \ln u}{dx} = \frac{1}{u} \frac{d u}{dx} = \Phi(x) ,  
       \label{4.dlnudx}
\end{equation} 
where 
\begin{equation} \label{eq:Phi_def}
\Phi(x)=\frac{1}{x-1}-\frac{3}{x}-\frac{d}{dx}\ln\!\bigl(H_1(x)H_2(x)\bigr).
\end{equation}
The master equation (\ref{4.masteqx}) may be rewritten as
 \begin{equation} \label{eq:Phi_0}
 \Phi(x)= 0.
 \end{equation}

 {\bf Proposition 1.} {\it For all $\mu > 0$, $P_1 > 0$ and  $P_2 > 0$ 
the  equation  (\ref{4.masteqR_0}) 
has one and only one real solution $R_0$ which satisfies the inequality $R_0 > 2 \mu$}.

 Proposition 1 states the existence and uniqueness of photon 
sphere (outside the event horizon) for the metric (\ref{i.3}) 
for any proper set of parameters. It is valid due to the following proposition.
 
{\bf Proposition 2.} {\it For all  $p_1 > 0$,  $p_2 > 0$ 
the fifth order polynomial (reduced master) equation  (\ref{4.masteqx})
has one and only one real solution $x = x_{0}$ which satisfies the inequality $x_{0} > 1$.} 

 {\bf Proof. }
 Let us set
  \begin{equation}
 z=\frac{1}{x},\qquad x>1 \iff z \in (0,1).
 \end{equation}
 We get
  \begin{equation}
 H_s(x)=1+\frac{p_s}{x}+\frac{q_s}{x^2}=1+p_sz+q_sz^2,
 \end{equation}
$s = 1,2.$
The equation $\Phi(x)=0$ from \eqref{eq:Phi_0} can be rewritten as
 \begin{equation}
 \label{eq:Phi_rewritten}
 \Phi\!\left(\frac1z\right)=0.
 \end{equation}

  A direct calculation yields
 \begin{equation}
 \label{4.Phiz}
  \Phi\!\left(\frac1z\right)
 = \frac{z\,Q(z)}{(z - 1)\, H_1(1/z)  \, H_2(1/z)},
 \end{equation}
 where
 \begin{eqnarray}
 \nonumber
  Q(z) = q_1q_2z^5-2q_1q_2z^4
    -\bigl(p_1p_2+p_1q_2+p_2q_1+q_1+q_2\bigr)z^3
 \\ -2(p_1+p_2)z^2 +(p_1+p_2-3)z+2. \quad  
  \end{eqnarray}
 Due to $z\in(0,1)$ we have
 \begin{equation} 
 z>0, \quad z-1<0, \quad H_s = 1+ p_sz + q_sz^2 > 0,
  \label{4.zH}
  \end{equation}
 $s = 1, 2$. 
 The  inequalities $H_s > 0$ in (\ref{4.zH}) just follow
 from
 \begin{equation} 
  p_s >0,\qquad q_s > 0,
   \label{4.psqs}
   \end{equation}
  $s = 1, 2$. 
Due to relations  (\ref{i4.qs}), (\ref{4.dudx}) and  (\ref{4.Phiz})
we obtain 
 \begin{equation} 
   z^5 F(1/z) = 4 (p_1 + p_2 + 2)^2 Q(z)
    \label{4.FQ}
    \end{equation}
 for $z \in (0,1)$. 
 
 Thus, the polynomial master equation (\ref{4.masteqx}) (or equivalently  $\Phi(x)=0$) for $x \in (1,+\infty)$ 
 is equivalent to another polynomial equation
 \begin{equation}
 Q(z)=0
 \label{4.masterQ}
 \end{equation}
 for  $ z \in (0,1)$ ($z = 1/x$).
 
 Let us prove that there exists at least one  solution to (\ref{4.masterQ})
 (or, to (\ref{4.masteqx}) ).
  We have
 \begin{equation}
  Q(0)=2>0, \label{4.Q0}
 \end{equation}
 while
 \begin{eqnarray}
 \nonumber
  Q(1) 
 = q_1q_2-2q_1q_2  -(p_1p_2+p_1q_2+p_2q_1+q_1+q_2)   
 -2(p_1+p_2)+(p_1+p_2-3)+2        \nonumber \\
 = -\bigl(p_1p_2+p_1q_2  + p_2q_1+q_1q_2      
  + p_1 +p_2 + q_1+ q_2 +1\bigr)<0. 
 \label{4.Q1} 
 \end{eqnarray}
 
   Function $Q(z)$ is continuous on $[0,1]$ and satisfies the boundary 
   conditions  (\ref{4.Q0}) and (\ref{4.Q1}). 
   The Intermediate Value (Bolzano) theorem implies
  the existence of at least one point $z_{\mathrm{0}}\in(0,1)$  which obeys
  \begin{equation}
  Q(z_{\mathrm{0}})=0. \label{4.Qz0}
  \end{equation}
   
\begin{figure}
  \includegraphics[width=\columnwidth]{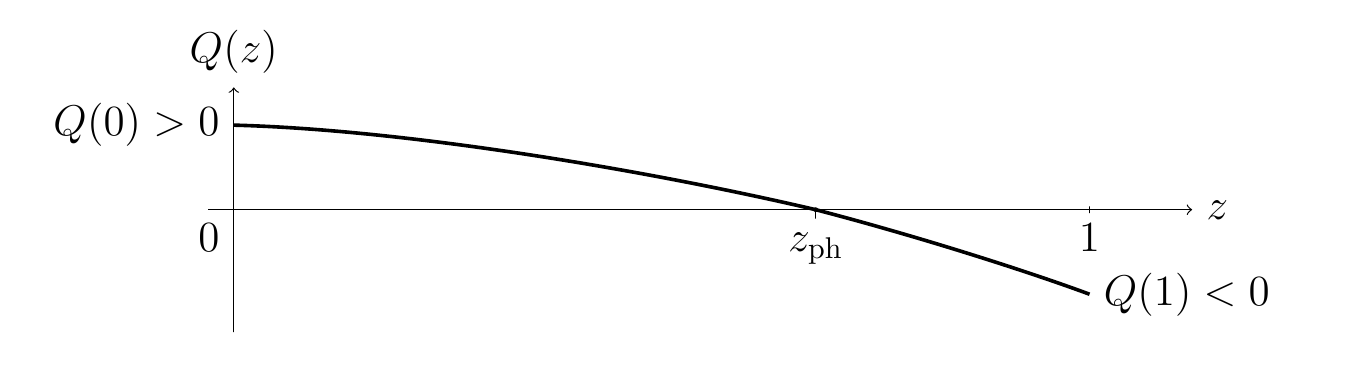}
  \caption{Schematic graph of the function $Q(z)$ on the interval $(0,1)$:
  $Q''(z)<0$, $Q(0)>0$, and $Q(1)<0$. Here $z_0 = z_{ph}$ corresponds to photonic sphere.}
  \label{fig:Q}
  \end{figure}

 Now compute the second derivative:
 \begin{eqnarray}
 Q''(z)
 =  20q_1q_2z^3-24q_1q_2z^2  - 6\bigl(p_1p_2+p_1q_2+p_2q_1+q_1+q_2\bigr)z 
 -4(p_1+p_2),  \label{4.d2Qdz2}
  \end{eqnarray}
 (Here and below $A' = \frac{dA}{dz}$, $A'' = \frac{d^2A}{dz^2}$.)
 Rewrite it as
 \begin{equation}
 Q''(z) =  4q_1q_2z^2(5z-6) 
 -6\bigl(p_1p_2+p_1q_2+p_2q_1+q_1+q_2\bigr)z 
 -4(p_1+p_2). \label{4.d2Qdz2a}
 \end{equation} 
 For $z \in (0,1)$, we have $5z-6<0$. Since all coefficients
 $p_s,q_s$ are positive  we obtain
 \begin{equation}
  Q''(z)<0, \qquad 
 \label{4.d2Qdzl0}
  \end{equation}
 for all $z \in (0,1)$.  See  Figure \ref{fig:Q}.

 Hence the function $Q$ is strictly concave on $(0,1)$, i.e.
   \begin{equation}
  Q\bigl((1-t)u_1+t u_2\bigr)   >  (1-t)Q(u_1)+tQ(u_2)
  \label{4.scQ}
   \end{equation}
 for all $t \in (0,1)$ and $0 < u_1 < u_2 < 1$.  
 This is the two-point case of Jensen's inequality for a concave function. 
 Geometrically, since $Q$ is strictly concave, 
 its graph lies strictly above the chord
 connecting the points $(u_1,Q(u_1))$ and $(u_2,Q(u_2))$.
 
 Now we prove the uniqueness of the root $z_0$.
 Assume the contrary: suppose there exists another 
 root $z_{1} \in (0,1)$, which obeys
  \begin{equation}
 Q(z_{1}) =0.
 \label{4.Qz1}
  \end{equation}
Without loss of generality we put $ z_{1} < z_0$.
Let $z_2 > 0 $ is small enough such that $0 < z_2  < z_{1}$ 
and  $Q(z_2) > 1$.
This is valid due to  $Q(z) \to Q(0) = 2$ for $z \to 0$ since 
the function $Q(z)$ is continuous on $[0,1]$.   
 Due to concavity relation (\ref{4.scQ}) we get
  \begin{equation}
   Q ((1-t) z_2 +t z_0)
   >  (1-t)Q(z_2)+tQ(z_0)
  \label{4.scQeps}
   \end{equation}
 for $t \in (0,1)$. Since $0 < z_2 < z_1 < z_0$
 there exists $t_1  \in (0,1) $ obeying 
  \begin{equation}
   (1-t_1)z_2 +t_1 z_0 = z_1. 
    \label{4.z1}
   \end{equation}
 Then,
 due to  (\ref{4.scQeps}) and $Q(z_0) = 0$ we obtain
the relation
 \begin{equation}
  Q (z_1)     >  (1-t_1)Q(z_2) > 0,
   \label{4.scQz1}
  \end{equation}
 which contradicts to the equality $Q (z_1) = 0$.  
  The obtained contradiction completes the proof
  of the Proposition 2. 
  
  {\bf Remark.} It should be noted that  the Proposition 2  
   was proved for general $q_s > 0$, $s = 1,2$, without using the relations (\ref{i4.qs}). 
   That means that the Proposition 1 is valid for a  class of black holes 
   (e.g. with  anisotropic  fluid matter source) described by metric (\ref{i.3}) 
   and moduli functions $H_s$ given by (\ref{i4.5}) 
   with arbitrary parameters $P_s > 0$ and $P_s^{(2)} > 0$, $s = 1,2$.

  {\bf Lemma 1.} {\it In notations of the proof of the Proposition 2 the following relation is valid: $Q'(z_0) < 0$.}
 
   {\bf Proof.} For strictly concave function we have well-known unequality
  \begin{equation}
                Q(z) < Q'(z_0)(z - z_0) + Q(z_0)
  \label{4.tang}
         \end{equation}
 for all $z \neq z_0$. Let us set   $z = z_2 < z_0$ with $2 > Q(z_2) > 1$.
 (Such real number does exist due to   $Q(0) = 2$, $Q(z_0) = 0$ and Intermediate Value theorem). 
 We obtain from (\ref{4.tang}) and   $Q(z_0) = 0$:
 \begin{equation}
                1 < Q(z_2) < Q'(z_0)(z_2 - z_0)
   \label{4.tangz2}
  \end{equation}
 which implies $Q'(z_0) < 0$.
  
 {\bf Proposition 3.} {\it For all $\mu > 0$, $P_1 > 0$,  $P_2 > 0$  
 the circular photonic geodesics given by unique solution  to master equation  
 (\ref{4.masteqx}) (or, equivalently $\left(\frac{d U}{d R}\right)_{R = R_0} = 0$)
  with $ R_0 = 2 \mu x_{0} > 2 \mu$  
 are unstable due to inequality 
 \begin{equation}
 \left(\frac{d^2 U}{d R^2}\right)_{R = R_0} < 0.
 \label{4.d2UdR2neg}
        \end{equation}
}

{\bf Proof.}
It follows from   (\ref{4.dUdR}),  (\ref{4.dudx})  and $R = 2 \mu x$ that
\begin{equation}           
      \left(\frac{d^2 U}{d R^2}\right)_{R = R_0} =   \frac{\hat{L}^2}{(2 \mu)^4 } 
      \left(\frac{d^2 u}{dx^2}\right)_{x = x_0},  
       \label{4.d2UdR2}
\end{equation} 
where 
\begin{equation}           
       \left(\frac{d^2 u}{dx^2}\right)_{x = x_0} = 
       -  \frac{ \left(\frac{d F}{dx}\right)_{x = x_0}}{4 (p_1 + p_2 + 2)^2 x^8(H_1 H_2)^2},  
       \label{4.d2udx2}
\end{equation} 
where $F(x)$ is given by (\ref{4.F}) and $x_0$ is the unique root of 
fifth order polynomial $F(x)$, i.e.
$F(x_0) = 0$, which obeys $x_0 >1$ (see Proposition 2).  

So, we need to prove only that 
\begin{equation}           
       \left(\frac{d F}{dx}\right)_{x = x_0} > 0.  
       \label{4.dFdx0}
\end{equation} 
Due to relation  (\ref{4.FQ}), $F(x_0) = 0$  and $x = 1/z$
\begin{equation} 
   z_0^5( -1/z^2_0)  \left(\frac{d F}{dx}\right)_{x = x_0} = 4 (p_1 + p_2 + 2)^2 Q'(z_0),
    \label{4.FQz0}
    \end{equation}
 where $ 0< z_0 = 1/x_0 <1$. But according to Lemma 1:  $Q'(z_0) < 0$. Hence the relation (\ref{4.dFdx0})
 is valid. Thus, the proposition is proved.

    \section{Examples}

     Here we consider certain examples of exact solutions to master equation 
     $F(x) = 0$ with fifth order polynomial $F(x)$ defined in (\ref{4.F}).      
      Some numerical roots of the reduced master equation are presented  in  Table \ref{tab:A2_roots}. 
   
   {\bf Example 1.}
   Let us  put 
   \begin{equation} 
   \label{4.P}
    p_1 = p_2 = p > 0.
   \end{equation}
   Then the dimensionless (reduced) master equation (\ref{4.masteqx}) reads
     \begin{equation} 
   \label{4.ZZ}
    (p+1)^2 (2x+p)^3(4x^2- 2px - 6x + p) =0.
       \end{equation}
  
   After excluding the   roots    $x_{1}= - p/2 < 0$,  
   we are led to   quadratic equation 
    \begin{equation} \label{4.7x}
      x^2 - \frac{1}{2} \left( p + 3  \right)  x   + \frac{1}{4} p = 0 
   \end{equation}
   which yields other two solutions   
      \begin{eqnarray}
   x_{\pm}&=&\frac{1}{4} \left( p + 3  \right)  \pm \frac{1}{4}\sqrt{d}, \label{4.8x}  \\
   d& =&   p^2 + 2 p  + 9> 0. \label{4.8d}
   \end{eqnarray}
     It could be readily verified that   
   \begin{equation} 
     \label{4.8z}
     x_3 = x_{+} > 1 > x_{-} = x_2 > 0
   \end{equation}
    for $p > 0$. 
         Thus, our unique solution $x_0 >1$ reads
     \begin{equation}
          \label{4.8x0}
         x_0 =  x_{+} = \frac{1}{4} \left( p + 3  \right)  + \frac{1}{4}\sqrt{ p^2 + 2 p  + 9}.
    \end{equation}
    
     In this case our metric is coinciding with  the
    Reissner-Nordstr\"om metric (\ref{i.RN_true}) written in
    radial variable $r = R +  \frac{P}{2}$. The relation (\ref{4.8x0}) for $x_0 = R_0/(2 \mu)$
    together with formulae (\ref{i.RN_M_Q})
    lead us to well-known relation for the radius of RN photon sphere $r_0 = R_0 +  \frac{P}{2}$
    in terms physical variables  $2GM$ and $Q^2_{RN}$:  
     \beq \label{i.RNr0}
     r_0 = \frac{1}{2} \left( 3 GM +  \sqrt{9(GM)^2 - 8Q^2_{RN}} \right). 
      \eeq  
         
    {\bf Example 2.} 
   For $p_1 = 2$, $p_2 = 1$, the given equation simplifies to  
  \begin{equation}
            \label{4.9p2q1}
        25x^5 - 75x^3 - 55x^2 - 6x + 3 = 0,   
      \end{equation}
      which factors as  
      \begin{equation}
      \label{4.9fact}
      (5x^2 + 5x - 1)(5x^3 - 5x^2 - 9x - 3) = 0.
     \end{equation}
   Hence the solutions are the roots of the quadratic and the cubic equations.
   The roots  of quadratic equation $5x^2 + 5x -1 = 0$  are real but less than $1$.  
   The qubic equation $5x^3 - 5x^2 - 9x - 3 = 0$ gives only one real root 
     \begin{equation} 
       \label{4.9qub1}
      x_0 = \frac{1}{3} + \sqrt[3]{\frac{430+18\sqrt{65}}{675}} + \sqrt[3]{\frac{430-18\sqrt{65}}{675}}
      \approx 2.031, 
      \end{equation}   
     which obeys $x > 1$. Two other roots are essentially complex.       
   
   {\bf Example 3.} Let us put $p_1 = k_1 p$, $p_2 =  k_2 p $, 
   where $k_1 > 0$, $k_2 > 0$ are fixed constants obeying 
   $k_1 + k_2 = 1$ and $p > 0 $. 
   We obtain two asymptotic relations for the root $x_0$  
     \begin{eqnarray}
      x_0 = \frac{3}{2} + \frac{p}{3} + O(p^2), \ {\rm as} \ p \to +0, \label{4.x3smallp}  \\
     x_0 \sim X p , \ {\rm as} \ p \to + \infty. \label{4.x3bigp}
      \end{eqnarray}
    
   In (\ref{4.x3bigp}) $X  > 0$ is a  solution to the 
   quartic equation 
      \begin{equation}
         \label{4.9Cquartic}
         P(X) = X^4 + \frac{1}{2}X^3 - \frac{1}{2} K^2 X  - \frac{1}{4} K^3 =0, 
         \end{equation}
         where $K = k_1 k_2$, $k_1 + k_2 =1$ ($k_1 > 0$, $k_2 > 0$). 
         It is evident that $0 < K = k_1 k_2 \leq \frac14$.
         The following proposition is valid.         
         
         {\bf Lemma 2.} {\it Only one root of quartic equation  (\ref{4.9Cquartic}) with $0 < K \leq \frac14$
          is positive.    It obeys  $0< X = X(K) \leq 1/4$.}
               
        The proof of the Lemma 2 will be given in a separate publication devoted to extremal case ($\mu = +0$) of the 
        black hole solution  under consideration. 
       
       {\bf Examples 4.} Certain numerical examples of roots of the reduced  master equation 
              (\ref{4.masteqx})  for several values of $p_1$ and $p_2$ are presented in Table  \ref{tab:A2_roots}.
  
       \begin{table}[ht]
       \centering
       \caption{Examples of roots of the reduced  master equation 
       (\ref{4.masteqx})  for several values of $p_1$ and $p_2$.}
       \label{tab:A2_roots}
       \begin{tabular}{c c c c c c c}
       \hline
       $p_1$ & $p_2$ & $x_1$ & $x_2$ & $x_3$ & $x_4$ & $x_5$ \\ 
        \hline
       $0.03$& $0.01$& $-0.019$ & $ -0.005 - 0.003i$ & $ -0.005 + 0.003i$ & $0.003$ & $1.507$ \\ 
       $0.1$ & $0.1$ & $-0.050$ & $-0.050$ & $-0.050$ & $0.016$ & $1.534$ \\ 
       $0.3$ & $0.1$ & $-0.187$ & $-0.053-0.027i$ & $-0.053+0.027i$ & $0.027$ & $1.565$ \\ 
       $1$ & $1$ & $-0.500$ & $-0.500$ & $-0.500$ & $0.134$ & $1.866$ \\ 
       $2$ & $1$ & $-1.171$ & $-0.516-0.171i$ & $-0.516+0.171i$ & $0.171$ & $2.031$ \\ 
       $2$ & $2$ & $-1.000$ & $-1.000$ & $-1.000$ & $0.219$ & $2.281$ \\ 
       $3$ & $1$ & $-1.760$ & $-0.541-0.229i$ & $-0.541+0.229i$ & $0.191$ & $2.152$ \\ 
       $3$ & $2$ & $-1.708$ & $-1.005-0.231i$ & $-1.005+0.231i$ & $0.246$ & $2.472$ \\ 
       $3$ & $3$ & $-1.500$ & $-1.500$ & $-1.500$ & $0.275$ & $2.725$ \\ 
       $4$ & $1$ & $-2.330$ & $-0.560-0.264i$ & $-0.560+0.264i$ & $0.204$ & $2.246$ \\ 
       $4$ & $2$ & $-2.313$ & $-1.037-0.320i$ & $-1.037+0.320i$ & $0.262$ & $2.624$ \\ 
       $4$ & $3$ & $-2.238$ & $-1.494-0.272i$ & $-1.494+0.272i$ & $0.294$ & $2.931$ \\ 
       $4$ & $4$ & $-2.000$ & $-2.000$ & $-2.000$ & $0.314$ & $3.186$ \\ 
       $5$ & $1$ & $-2.886$ & $-0.574-0.287i$ & $-0.574+0.287i$ & $0.213$ & $2.320$ \\ 
       $5$ & $2$ & $-2.897$ & $-1.063-0.378i$ & $-1.063+0.378i$ & $0.274$ & $2.749$ \\ 
       $5$ & $3$ & $-2.856$ & $-1.527-0.386i$ & $-1.527+0.386i$ & $0.307$ & $3.104$ \\ 
       $5$ & $4$ & $-2.763$ & $-1.983-0.304i$ & $-1.983+0.304i$ & $0.328$ & $3.402$ \\ 
       $5$ & $5$ & $-2.500$ & $-2.500       $ & $-2.500       $ & $0.342$ & $3.658$ \\ 
       $10$& $5$& $-5.722$ & $-2.601-0.740i$ & $-2.601+0.740i$ & $0.375$ & $4.548$ \\
       $10$& $10$&$-5.000$ & $-5.000       $ & $-5.000       $ & $0.411$ & $6.089$ \\
       \hline
       \end{tabular}%
       \end{table}
       

     \section{BH shadow} 
       
      Following the standard procedure \cite{PTs,IBBKMNZ}, one arrives at a particular solution which takes the form of a geodesic with a spiral-like shape and an infinitely large winding angle. This trajectory acts as a separatrix between the two distinct families of solutions—those that intersect the photon sphere and those that do not. It directly corresponds to the critical shadow angle $\vartheta_{sh}\in (0, \pi/2)$, which is geometrically determined at the observer's position $(R_{obs}, \phi_{obs})$ as the angle formed by the tangent to this spiral curve and the local radial direction.

      Let a light ray be emitted from $(R_{obs}, \phi_{obs})$ at an angle smaller than the critical angle:
      \begin{equation}\label{b.104}
      \vartheta < \vartheta_{sh},
      \end{equation}
      then it will traverse the photon sphere and  enter the event horizon (and ``fall'' into the black hole).
      In opposite case, when the emission angle obeyss
      \begin{equation}\label{b.105}
      \vartheta > \vartheta_{sh},
      \end{equation}
      the ray will not reach the photon sphere and will escape to infinity 
      (after a finite number of revolutions).    
        
         Let us define reduced effective potential as 
           \beq \label{4.eff_pot_2}
         \hat{U}(R) = U/\hat{L}^2 =  \frac{A(R)}{ C(R) }   \qquad  \\ \nonumber
            = (H_1 H_2)^{-1} \left(1-\frac{2\mu}{R}\right)
             \frac{1}{ R^2},
          \eeq  
        where $\hat{L} \neq 0$.                  
        $R_0$ corresponds to the point of maximum of the effective potential and hence
      \begin{equation}\label{b.107}
      \hat{U}(R_0) > \hat{U}(R) > 0
      \end{equation}
       for  $R > R_0$.
       
       The black hole shadow angle $\vartheta_{sh}$ can be found 
      by using a well-known relation \cite{PTs} (see also \cite{IBBKMNZ}):
      \begin{equation}\label{b.114}
      \vartheta_{sh} = \arcsin\sqrt{\frac{\hat{U}(R_{obs})}{\hat{U}(R_0)}}, 
      \qquad 0< \vartheta_{sh}< \frac{\pi}{2},
      \end{equation}
      for all $R_{obs} > R_0$.  
      Here, $R_{obs}$ denotes the radial coordinate  of a point-like source or 
      light receiver (e.g. observer), while $R_0$ represents the radius of the photon sphere.

      For sufficiently large values of the observer's radial coordinate $R_{obs}$
      satisfying $R_{obs} \gg P$ and $R_{obs} \gg \mu$, the shadow angle can be found
      by using the following asymptotic formula:
      \begin{equation}\label{b.115}
         \vartheta_{sh} = \frac{b_c}{R_{obs} } + O ( 1/R^2_{obs}), 
      \end{equation}
      as $R_{obs} \to + \infty$. 
       Here
       \begin{equation}\label{b.116}
         b_c  = \frac{1}{\sqrt{\hat{U}(R_0)}}.
         \end{equation} 
      is critical impact parameter \cite{PTs}. 
      
       For dyonic BH under consideration    we obtain
        \begin{equation}
         b_{c} = R_0 ( H_1(R_0) H_2(R_0))^{1/2} \left(1-\frac{2\mu}{R_0}\right)^{-1/2}.
       \label{b0}
       \end{equation}
      It is evident, that $ b_{c} > R_0$ for all $\mu > 0$, $P_1 > 0$ and $P_2 > 0$.
     
      Let us fix $\mu > 0$. We get 
      \begin{equation}
      \label{b.118}
       b_{c} \to  3 \sqrt{3} \mu 
      \end{equation}
     in the (Schwarzschild) limit when $P_1 \to +0$, $P_2 \to +0$.
     
      For  $P_1 \to + \infty$, $P_2 \to + \infty$ and $P_1/P_2 = k > 0$ 
      ($k$ is constant)   we are led to the asymptotical relation 
            \begin{equation}
      \label{b.119}
            b_c \sim 2 \mu X p ( H_{1,\infty} H_{2, \infty})^{1/2} 
      \end{equation}
       (see (\ref{4.x3bigp})) as $p = (P_1 + P_2)/(2 \mu) \to + \infty$,
       where
            \begin{equation}
                H_{1,\infty}= 1+\frac{K_1}{X}  +\frac{K_1^2 K_2}{2 X^2},
                  \quad        H_{2,\infty}=1+\frac{K_2}{X} +\frac{K_1 K_2^2}{2 X^2}.
                \label{Hinf}
            \end{equation}
      Here $K_s = P_s/(P_1 + P_2) > 0$, $s = 1,2$, are constants and $X = X(K_1 K_2) > 0$
      is a solution to quartic equation  (\ref{4.9Cquartic}). The relation (\ref{b.119}) just follows
      from $x_0 = R_{0}/(2 \mu) \sim X p$, as $p \to + \infty$ and relations for moduli functions $H_s$, $s = 1,2$,
      from (\ref{i4.5}). 
      
       Now, let us put $P_1 = P_2 = P$. We obtain $R_0 = 2 \mu x_0$,                   
                 $H_1(R_0) = H_2(R_0) = \left(1 + \frac{P}{2R_0} \right)^2$ and hence 
                 \begin{equation}
                 \label{b0_sym}
                            b_c = 2 \mu x_0  \left(1 + \frac{p}{2x_0} \right)^2 \left(1-\frac{1}{x_0}\right)^{-1/2}, 
                 \end{equation}
    where $p = P/(2 \mu)$ and $x_0 =   \frac{1}{4} \left( p + 3  \right)  + \frac{1}{4}\sqrt{ p^2 + 2 p  + 9}$ (see (\ref{4.8x})). Relation (\ref{b0_sym}) is in an agreement with the well-known formula for critical impact parameter of the  non-extremal  Reissner-Nordstr\"om  charged black hole \cite{Vert}:
                \begin{equation}
                    \label{b0_RN}
                  b_c = \frac{r_0}{\sqrt{A(r_0)}}, 
                 \end{equation}
                 where $r_0 = R_0 +  \frac{P}{2} = R_0 ( H_1(R_0) H_2(R_0))^{1/4}$ and $A(r_0) =  1 - \frac{2GM}{r_0} +  \frac{Q^2_{RN}}{r_0^2}$.

      Relation (\ref{b0}) implies the identity for the photon capture cross section
               \begin{equation}
             \sigma_{cap} = \pi b^2_{c} = R^2_0 ( H_1(R_0) H_2(R_0)) \left(1-\frac{2\mu}{R_0}\right)^{-1}.
              \label{sigma_cap}
        \end{equation}
      For $P_1 = P_2 = P$ we are lead to the  photon capture cross section of 
      the  non-extremal  Reissner-Nordstr\"om   black hole
        $\sigma_{cap} = \frac{\pi r_0^2}{A(r_0)}$.
         
\section{Conclusions}
  
   In this work we have studied the non-extremal dyonic black hole generated by the $A_2$ Toda chain 
  for a dilatonic coupling parameter   $\lambda$ obeying  $\lambda^2=3/2$. 
  The associated moduli functions are quadratic polynomials in inverse radial variable $1/R$, 
  and the condition for a circular null geodesic yields a fifth-order polynomial master equation for the radius of
   photon sphere $R_0$. We have proved that this polynomial possesses exactly one real root outside the horizon
    \(R_0>2\mu\) and that  the corresponding photon orbits are unstable. The proof uses only elementary properties of a strictly concave   auxiliary polynomial and does not rely on energy conditions.

   We note that the existence of a unique unstable photon sphere for this solution is already implied by the result of 
   Cveti\v{c}, Gibbons and Pope \cite{CGP}. The new results here are the explicit fifth-order polynomial master equation, the self-contained concavity argument for the proof of the uniqueness  
    of the photon sphere outside the horizon, the calculation 
   of  the second derivative of the effective potential $U''(R_0)$, the examples of analytical and numerical roots, 
   an explicit expression for the shadow angle, the critical impact parameter 
   and the capture cross-section. The complementary case with $\lambda^2=1/2$ and cubic master equation was treated 
   in Ref. \cite{IKMN}; the two analyses are complementary rather than overlapping.
   
  The results may be used in future studies of quasinormal modes in the eikonal limit and of innermost stable circular orbits.

	\small
	\bibliographystyle{scilight}
	
	

\end{document}